# Spectrum Sharing between Directional-Antenna-Equipped UAV System and Terrestrial Systems

Tao Yu, Kento Kajiwara, Kiyomichi Araki, Kei Sakaguchi
Tokyo Institute of Technology, Japan
{yutao, kajiwara, araki, sakaguchi}@mobile.ee.titech.ac.jp

*Abstract*—Unmanned aerial vehicles (UAVs)-based applications, such as surveillance systems and wireless relays, are attracting increasing attention from academia and industrial fields. The high-performance aerial communication system is one of the key enablers for them. However, due to the low attenuation of radio waves in the air-to-ground channels, the interference between aerial and terrestrial communication systems would significantly deteriorate their communication performance and greatly limit the potential UAV applications. To address the problem, in this paper, the spectrum sharing strategy between a multiple UAV communication system, in which both UAVs and ground station (GS) are equipped with directional antennas, and terrestrial systems is proposed. The GS position is selected and the flyable areas of the UAVs using certain spectrum resources are defined in advance using prior knowledge from spectrum monitoring on terrestrial communication systems to minimize interference and maximize the flyable areas of the UAVs instead of the low-efficient dynamic channel sensing and allocation for interference elimination. The simulations are conducted through a case study of the spectrum sharing between a multi-UAV video transmission system and the terrestrial wireless local area network (WLAN) system in the 5.7GHz band. The simulation results show that thanks to the proposed system the entire area can be enabled for UAV flight.

*Keywords—UAV communication, Spectrum Sharing, directional antenna*

I. INTRODUCTION

The labor shortages and growing labor expenses have been a concern in aging populations with falling birthrates in recent years. As a result, applications of unmanned systems using robots, such as unmanned aerial vehicles (UAVs), to cut costs are attracting increasing attentions. The market for UAVs and their related services is quickly developing, with a market value of $1.3 trillion expected by 2040[1]. UAVs are expected to be employed in applications such as disaster relief, logistics, agricultural, infrastructure inspection, and security monitoring [2][3].

The high-performance communication is one of the fundamental technologies for realizing UAV systems and their associated applications because it enables real-time information sharing between UAVs and ground stations (GSs) via wireless communication and closes the control-sensing loop. However, various challenges continue to finder the development of high-performance multi-UAV communication systems, such as the inter-system interference and spectrum sharing strategy. Due to the lack of natural or artificial obstacles (e.g., hills and buildings) at the working height of UAVs, line-of-sight (LOS) paths are typically very likely to exist in the air-to-ground channels, which leads to severe co-channel interference from aerial and terrestrial wireless communication systems to each other. This does not only result in decreased system performance, but also in additional system costs associated with, e.g., the carrier sensing before transmission.

Currently, the majority of commercial aerial communication systems are employing solutions based on the standardized technologies such as IEEE 802.11/802.15 families, cellular families and their derivatives [4]-[6], so the carrier sensing multiple access with collision avoidance (CSMA/CA) has conventionally been widely used for sharing the same spectrum among different wireless user terminals (UEs) and different wireless systems. However, it would result in very low efficiency in the air-to-ground propagation environment due to the low attenuation of interference and long propagation delays. The transmit power and UAV's placement/trajectory can also be jointly optimized to achieve the spectrum sharing between UAVs and terrestrial device-to-device (D2D) communication systems and cellular networks [7][8]. In [9], a machine learning-assisted framework for spectrum sharing between UAV and ground D2D was developed, in which UAVs opportunistically access the licensed channels through spatial spectrum sensing.

In this paper, we propose a method to share the spectrum and reduce the interference between aerial and terrestrial communication systems using same spectrum resources by installing directional antennas on the UAVs and the GS to perform spatial and directional separation between aerial and terrestrial communication systems. The GS position is selected and the flyable areas of the UAVs using the same spectrum resources with terrestrial systems are defined in advance by using prior knowledge from spectrum monitoring on terrestrial systems, instead of the dynamic channel sensing and allocation. The proposed method aims to minimize negative effects on system performance of different systems and maximize the flyable areas of the UAVs. For the multiple UAV systems, the channel allocation scheme is also proposed to establish spectrum sharing conditions and to expand the area where UAVs can fly within the target area. Numerical simulations are also conducted by a case study of the multi-UAV full-duplex video transmission system proposed in our earlier works [10][11].

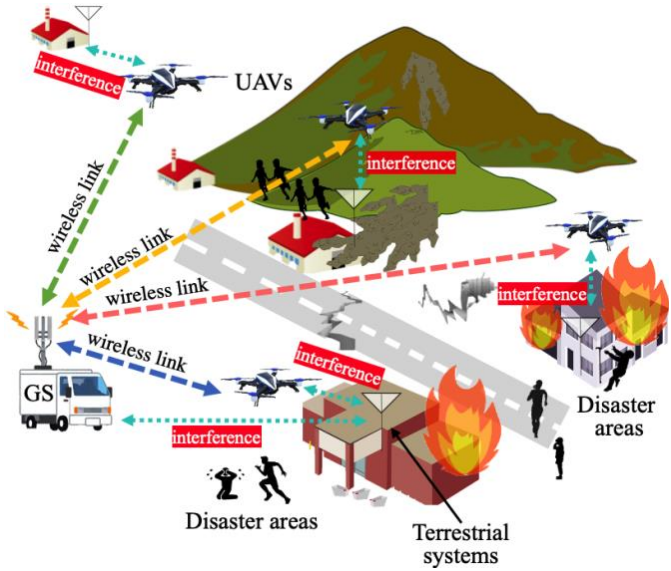

Fig. 1 Aerial and terrestrial communication systems

The rest of this paper is organized as follows. Sect. II presents the proposed system model and analysis conditions for spectrum sharing. Sect. III numerically analyzes the flyable area of UAVs without inter-system interference. Sect. IV discusses the channel allocation for multiple UAVs. Finally, Sect. V concludes the paper.

## II. SYSTEM ARCHITECTURE

An application scenario of disaster relief by a multi-UAV system is illustrated in Fig.1. Multiple UAVs take high-resolution videos in disaster areas and transmit them via wireless links to the GS carried on a vehicle, and in the meantime, the GS sends control flight control commands to UAVs. But in the disaster areas and nearby areas, many terrestrial communications systems exist, such as wireless local area network (WLAN), cellular network, and dedicated short-range communication (DSRC). UAV systems and terrestrial systems could have strong interference with each other if they use the same radio resources. The interference is bidirectional: the performance of UEs for both aerial and terrestrial systems could be deteriorated or even be disabled. Therefore, the spectrum sharing strategy must be introduced to avoid such an issue.

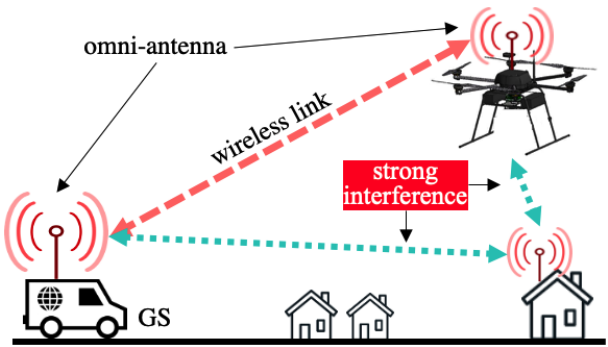

Fig. 2 Inter-system interference in conventional system

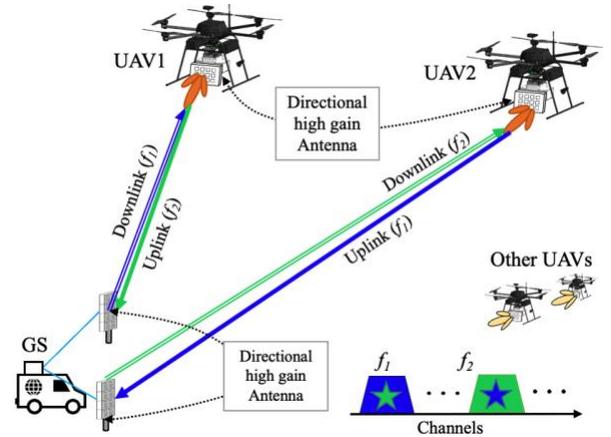

(a) Multi-UAV full-duplex communication system

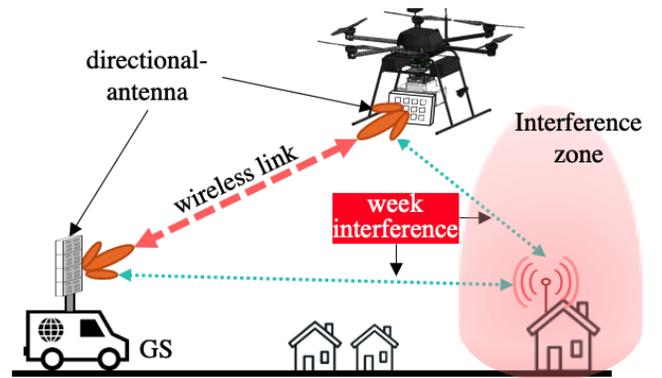

(b) Inter-system interference

Fig. 3 Inter-system interference in proposed system

### A. Conventional system architecture

Typically, terrestrial UEs are typically equipped with omni-antennas. Fig. 2 illustrates the interference between terrestrial systems and conventional UAV systems which also equip omni-antennas. Because of the omni-direction and high transmission power of UAV, severe interference between the two systems (i.e., interference between terrestrial UEs and GS, and interference between terrestrial UEs and UAV) would occur. Conventionally, the mechanism of carrier sense and collision avoidance is employed on both UAV systems and terrestrial systems, which results in that both systems can only utilize of part of radio resources and cannot perform continuous transmission. This issue would be magnified when UAV is performing tasks such as video surveillance for disaster relief in which high-resolution video needs to be continuously transmitted from UAV to GS.

### B. Proposed system architecture

In our previous works [10][11], a multi-UAV full-duplex system proposed was proposed, as shown in Fig.3(a). In the system, the uplink and downlink of a UAV employ two different and sufficiently separated channels, so that complicated hardware for self-interference cancellation in conventional IBFD architectures on a single UAV can be avoided, and the adjacent-channel interference between Tx/Rx can also be effectively suppressed by analog filters. To achieve high spectrum efficiency and realize multi-UAV IBFD, each channel is re-allocated to uplink of one UAV and downlink of another. All channels in the system are simultaneously reused by uplinks and downlinks of multiple UAVs.

Moreover, the proposed architecture cannot only reduce the intra-system interference between UAVs, but also can effectively eliminate the inter-system between the UAVs and terrestrial systems, as shown in Fig.3(b) compared with Fig.2. By taking the advantage of directional antennas, if the relative position of UAVs and terrestrial UEs, which use the same channels, can be well controlled to make the side lope of UAV antennas direct to terrestrial UEs, the intra-system interference can be well suppressed without the extra system costs such as carrier sensing, and the continuous transmission of both systems can be also realized.

First, a quick spectrum monitoring should be performed in the target area to get the distribution of terrestrial UEs that use the same band with UAV systems. It is noted that the distribution of these radio emitters is assumed to be not dynamic within time from monitoring to UAV operation. For example, the 5.7GHz band is shared by UAV and terrestrial system such as WLAN while the distribution of WLAN routers can be assumed to be static. Second, with the prior knowledge of terrestrial UE distributions and with the considerations of directivity of UAV antennas, UAVs' flyable areas, where the UAV can be operated, and interference aeras, where intra-system interference would occur, as shown in Fig.3(b), should be defined and the position of GS should be selected in advance of the UAV operation. Then, according to the flyable areas and interference areas, UAVs will be operated to perform tasks such as surveillance and disaster relief. In this paper, we assume that the first step has been done, i.e., the terrestrial UEs information is available, so we mainly focus on the second step, i.e., the derivation of flyable areas and interference areas.

C. *Spectrum sharing condistions*

The flyable areas in which the spectrum can be safely shared between UAVs and terrestrial systems is defined as the areas satisfying the following conditions, in which the deterministic model is used in this paper.

$$\begin{aligned} SINR_{\text{UAV}}^{\text{Up}} &> SINR_{\text{UAV,REQ}}^{\text{Up}} \\ SINR_{\text{UAV}}^{\text{Down}} &> SINR_{\text{UAV,REQ}}^{\text{Down}} \\ SINR_{\text{Terr}}^{\text{Up}} &> SINR_{\text{Terr,REQ}}^{\text{Up}} \\ SINR_{\text{Terr}}^{\text{Down}} &> SINR_{\text{Terr,REQ}}^{\text{Down}} \end{aligned} \quad (1)$$

where $SINR_{\text{UAV}}^{\text{Up}}$, $SINR_{\text{UAV}}^{\text{Down}}$, $SINR_{\text{Terr}}^{\text{Up}}$ and $SINR_{\text{Terr}}^{\text{Down}}$ are the signal-to-interference-and-noise-ratios (SINRs), respectively. $SINR_{\text{UAV,REQ}}^{\text{Up}}$, $SINR_{\text{UAV,REQ}}^{\text{Down}}$, $SINR_{\text{Terr,REQ}}^{\text{Up}}$ and $SINR_{\text{Terr,REQ}}^{\text{Down}}$ are the minimum required SINR by the UAV and terrestrial UEs for required communication service.

III. FLYABLE AREAS FOR SPECTRUM SHARING

The spectrum sharing between the terrestrial WLAN system and a multi-UAV full-duplex video transmission system, which was proposed in our earlier works [10][11], at 5.7 GHz band will be used as a case study in this section to show the effeteness of the proposed method.

A. *System model for spectrum sharing between UAVs and WLAN*

WLAN routers are considered as terrestrial UEs and are assumed to be equipped with omni-directional antennas. To evaluate the performance in the worst case, the LOS paths are assumed to exist between UAVs and routers, i.e., routers are assumed to be planced by the windows and the building entry losses are assumed to be 0 dB. The WLAN UEs are assumed to be less than 5 m away from the routers. UAVs use directional antennas, and the beam alignment of the directional antennas of the UAV and the GS is assumed to be perfect, so that the antennas in UAVs and GS are always direct to each other.

Several interferences exist in this model: 1) the interference from UAV to routers, 2) the interference from routers to UAV, 3) the interference from GS to routers, and 4) the interference from routers to GS, as shown in Fig.4. The log-distance path-loss model and empirical path-loss exponents are used. In this section, the air-to-ground channels (i.e., UAV to GS, UAV to routers) employ path-loss exponents of 2, and the ground-to-ground channels (i.e., GS to routers) employ path-loss exponents of 4.

The full-duplex multi-UAV communication system, which was proposed in our early works [10][11] for 4K video transmission for disaster relief, is taken as a representative example of UAV applications in this section. It uses frequency band from 5650MHz to 5750MHz and must share the spectrum from 5650MHz to 5730MHz with the WLAN on the ground [12], as shown in Fig.5. To support the 4K transmission and flight control, it requires minimum uplink SINR of 11 dB and minimum downlink SINR of 2 dB. Other parameters can be found in Table I.

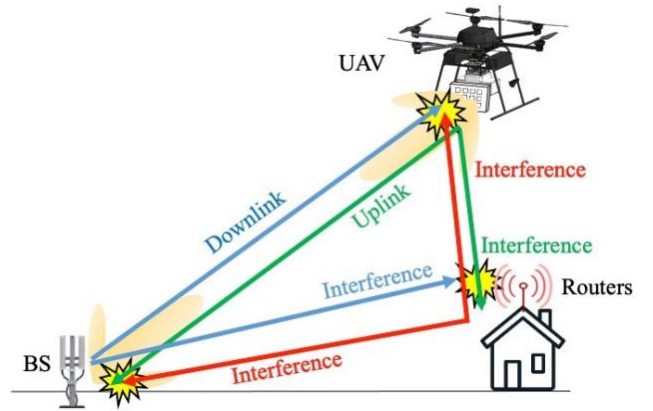

Fig.4 Interferences in the system

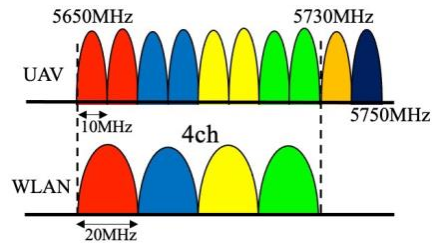

Fig.5 Spectrum usage of the UAV and WLAN system

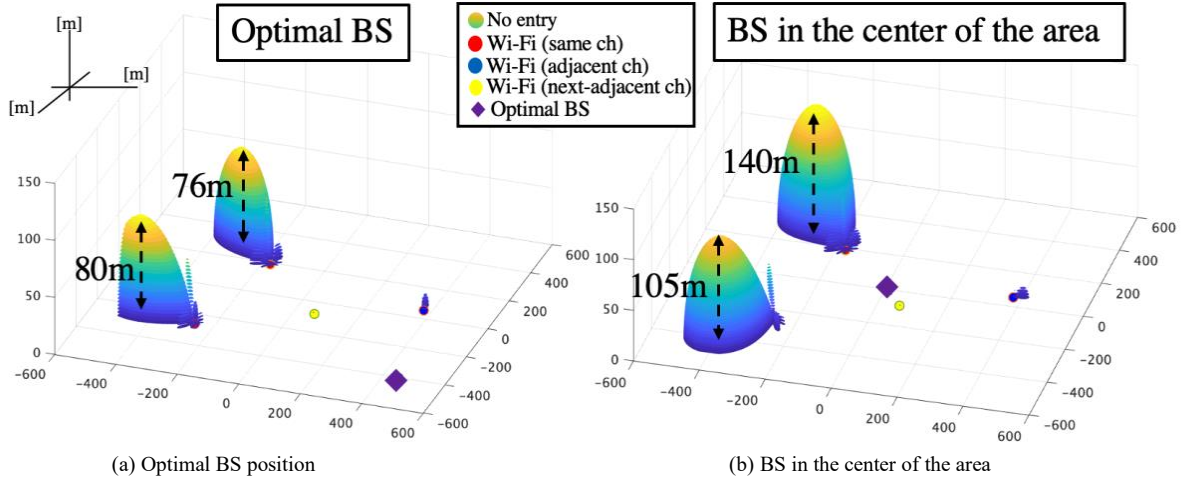

(a) Optimal BS position  (b) BS in the center of the area

Fig.6 UAV flyable areas

Table I  System Parameters

| Parameters | Values |
| --- | --- |
| Frequency Band | 5.7GHz |
| Bandwidth (down/uplink) | 10MHz/10MHz |
| Tx Power (GS/UAV) | 11dBm/0dBm |
| Antenna Gain (GS/UAV) | 25dBi/15dBi |
| Beamwidth (GS/UAV) | 4° / 36° |

The routers follow the standard IEEE802.11ac. According to the table of modulation and coding scheme (MCS), MCS-0 is selected as the minimum requirements of WLAN system for communication. The WLAN routers and users are close to each other compared with their distance to UAV and GS and are assumed to have same transmission power, so the interference to them from UAV and GS are approximately equal. Therefore, the following condition for spectrum sharing can be derived.

$$SINR_{UAV}^{Up} > 11 \text{ dB}$$
$$SINR_{UAV}^{Down} > 2 \text{ dB} \qquad (2)$$
$$SINR_{Terr} > 2 \text{ dB}$$

B. *Flyable Areas for Spectrum Sharing*

In the simulation, four WLAN routers (two of which use the same channel as the UAV and GS, and the other two use adjacent and next-adjacent channels) are randomly installed within area of 1 km$^2$. (Based on the IEEE802.11ac spectrum mask standard, the SNR required for channel rejection when the UAV and GS use the Wi-Fi adjacent channel is 16 dB, and the next adjacent channel rejection is 32 dB.) Assuming that the information of routers is known from the spectrum monitoring, the flyable area can be calculated, and the optimal GS position can be selected by an exhaustive method. The UAV is at 30m height.

The results are shown Fig.6(a). For comparisons, Fig. 6(b) shows the results for the case where the GS is placed in the center of the target area. The area colored in the figures are interference zones in which the condition in Eq.(2) cannot be satisfied. The result also show that interference zone can be reduced by selecting the optimal GS.

The proposed method and the conventional method are also compared in terms of the flyable area. In the conventional method, omni-directional antennas are employed in UAV and GS, i.e., antenna gain is 0 dB. The EIRP is limited to 36 dBm because of radio laws in Japan. The simulation altitude is 30m and the results are summarized in Fig. 7. The conventional method with omni-directional antenna has 0% of the flyable area when UAV is at 30m altitude, while the proposed method has 96% of the flyable area. Because the conventional method requires a much larger transmit power, the interference to WLAN routers becomes larger and the spectrum sharing condition Eq.(2) cannot be satisfied. In addition, the interference caused by routers becomes larger than the proposed method due to the omni-directional antenna, and the spectrum sharing conditions cannot be satisfied. From the above, it is considered that the proposed method reduces the interference with terrestrial WLAN systems by narrowing the beam and leads to the expansion of the UAV's flyable area. It is also shown that the UAV can expand its flyable area by selecting the position of the GS.

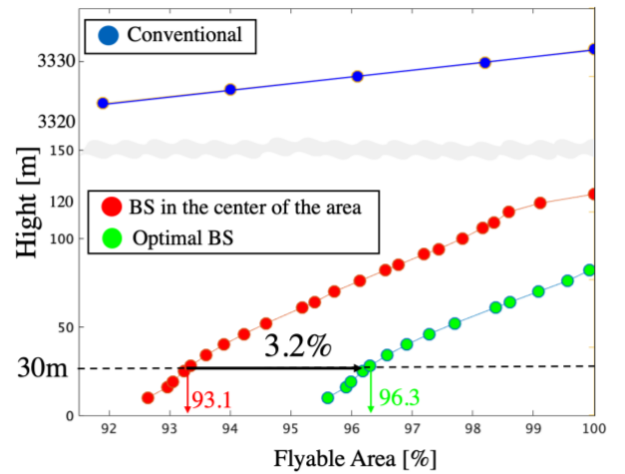

Fig. 7  Flyable area ratios comparison

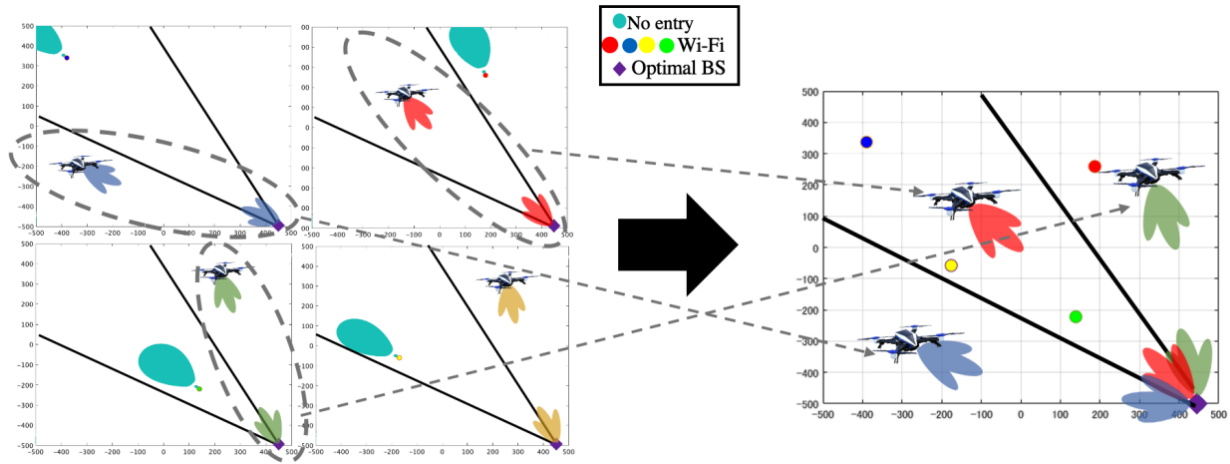

Fig. 8 Extended flyable area by channel allocation

C. *Channel allocation and Extend flight area*

Even after optimizing the location of the GS, the flyable area of the UAV does not cover the entire target area. Therefore, we devised a channel allocation method for UAVs to reduce interference. The whole targe area is divided into sub-areas for different UAVs using certain channels. The sub-areas are with equal area with the GS at the center. If we consider the case of operating with one UAV without dividing the area, the area that can be flown is 94% of the total area. On the other hand, as shown in Fig. 8, when the area was divided and operated with three UAVs, the entire area was flyable. Thirty different routers arrangements of four UAVs were used to derive the flyable area, and when three UAVs were used, the entire area was flyable.

## IV. CONCLUSION

In this paper, the spectrum sharing strategy between a multiple UAV communication system, in which both UAVs and GS are equipped with directional antennas, and terrestrial systems is proposed. The GS position is selected and the flyable areas of the UAVs using certain spectrum resources are defined in advance using prior knowledge from spectrum monitoring on terrestrial communication systems to minimize interference and maximize the flyable areas of the UAVs instead of the low-efficient dynamic channel sensing and allocation for interference elimination. The simulations are conducted through a case study of the spectrum sharing between a multi-UAV video transmission system and the terrestrial WLAN system in the 5.7GHz band. Simulation results confirm the effectiveness of the proposed system, and show that compared with conventional system, the proposed system can increase the flyable area by 96% when UAV is at 30m height. Moreover, by dividing the target area equally into the number of UAVs and assigning the channels other than those used by Wi-Fi in each area to the UAVs, we expanded the flyable area to the entire area.

ACKNOWLEDGMENT

The research leading to these results is funded by the Ministry of Internal affairs and Communications (MIC) of Japan under the grant agreements 0155-0083.

REFERENCES

[1] Goldman Sachs, "Drones Reporting Work", Goldman Sachs, https://www.goldmansachs.com/insights/technology-driving-innovation/drones/
[2] N. Hossein Motlagh, et al., "Low-Altitude Unmanned Aerial Vehicles-Based Internet of Things Services: Comprehensive Survey and Future Perspectives", IEEE Internet Things J., vol. 3, no. 6, pp. 899-922, 2016.
[3] Y. Zeng, et al., "Wireless communications with unmanned aerial vehicles: opportunities and challenges", IEEE Commun. Mag., vol. 54, no. 5, pp. 36-42, May 2016.
[4] S. Hayat, E. Yanmaz, and R. Muzaffar, "Survey on unmanned aerial vehicle networks for civil applications: A communications viewpoint." IEEE Commun. Surv. Tutor., vol.18.4, pp.2624-2661, 2016.
[5] "Enhanced LTE support for aerial vehicles", 3GPP TR 36.777, 2019.
[6] L. Bertizzolo, T. X. Tran, B. Amento, B. Balasubramanian, R. Jana,H. Purdy, Y. Zhou, T. Melodia, "Live and let Live: Flying UAVs Without Affecting Terrestrial UEs", ACM HotMobile, pp.21-26, 2020.
[7] H. Wang, et al., "Spectrum Sharing Planning for Full-Duplex UAV Relaying Systems With Underlaid D2D Communications," IEEE J. Sel. Areas Commun., vol. 36, no. 9, pp. 1986-1999, Sept. 2018.
[8] Y. Guo, et al., "Joint Placement and Resources Optimization for Multi-User UAV-Relaying Systems With Underlaid Cellular Networks," IEEE Trans. Veh. Technol., vol. 69, no. 10, pp. 12374-12377, Oct. 2020.
[9] B. Shang, L. Liu, R. M. Rao, V. Marojevic and J. H. Reed, "3D Spectrum Sharing for Hybrid D2D and UAV Networks," IEEE Trans. Commun., vol. 68, no. 9, pp. 5375-5389, Sept. 2020.
[10] T. Yu, S. Imada, K. Araki and K. Sakaguchi, "Multi-UAV Full-Duplex Communication Systems for Joint Video Transmission and Flight Control," IEEE CCWC, Jan. 2021.
[11] T. Yu, K. Araki and K. Sakaguchi, "Full-Duplex Aerial Communication System for Multiple UAVs with Directional Antennas", IEEE CCNC, 2022.
[12] Ministry of Internal Affairs and Communications, "The Radio Use Web Site ", https://www.tele.soumu.go.jp/j/sys/others/drone/.